\documentclass{article}
\usepackage[a4paper, total={6in, 8in}]{geometry}
\usepackage[utf8]{inputenc}
\usepackage{graphicx}
\usepackage{dirtytalk}
\usepackage[T1]{fontenc}

\title{\textbf{Simplified single-shot supercontinuum spectral interferometry} \\ \textit{\normalsize Institute for Research in Electronics and Applied Physics, University of Maryland, MD 20742, USA}}
\author{\normalsize D. Patel, D. Jang, S. W. Hancock, H. M. Milchberg, and K. Y. Kim}
\date{}

\begin{document}

\maketitle

\begin{abstract}
We have experimentally demonstrated a simplified method for performing single-shot supercontinuum spectral interferometry (SSSI) that does not require pre-characterization of the probe pulse. The method, originally proposed by D. T. Vu, D. Jang, and K. Y. Kim, uses a genetic algorithm (GA) and as few as two time-delayed pump-probe shots to retrieve the pump-induced phase shift on the probe [Opt. Express \textbf{26}, 20572 (2018)]. We show that the GA is able to successfully retrieve the transient modulations on the probe, and that the error in the retrieved modulation decreases dramatically with the number of shots used. In addition, we propose and demonstrate a practical method that allows SSSI to be done with a single pump-probe shot (again, without the need for pre-characterization of the probe). This simplified method can prove to be immensely useful when performing SSSI with a low-repetition-rate laser source.    
\end{abstract}

\section*{\normalsize Introduction}
Single-shot supercontinuum spectral interferometry (SSSI) is a spectral interferometric method that uses supercontinuum (SC) chirped laser pulses to probe a pump-induced ultrafast refractive index transient \cite{Kim:01}. Each spectral component of the chirped probe pulse encodes a time-resolved piece of information in the form of the pump-induced phase shift via cross-phase modulation. Interfering this probe pulse with a replica reference pulse in an imaging spectrometer yields a 2-dimensional (2D) (1D space and wavelength) interferogram from which the pump-induced phase shift (hence, the ultrafast transient) can be extracted . This technique, and variants \cite{Blanc,Geindre}, have been used to measure various phenomena including double-step ionization of helium \cite{Kim:02}, laser wakefields \cite{Matlis:03}, \(n_2\) measurement of air at mid- and long wave- IR wavelengths \cite{Zahedpour:04}, bound-electron nonlinearities near the ionization threshold for various gases \cite{Wahlstrand:05}, ultrafast nonlinear electronic, rotational, and vibrational responses in molecular gases \cite{Wahlstrand:06}, optical conductivity of laser-heated aluminum plasma \cite{Churina}, time domain terahertz waveform \cite{Teo}, polymorphic phase transitions in iron \cite{Crowhurst2}, and characterize laser-induced shock in materials \cite{Moore}.

Since this technique encodes temporal information onto various spectral components of a probe, the spectral phase of the probe pulse must be known to extract the time-resolved nonlinearity. Self-referencing diagnostics such as frequency-resolved optical gating (FROG) \cite{Trebino:07} or spectral phase interferometry for direct electric-field reconstruction (SPIDER) \cite{Iaconis:08} which are often used to characterize laser pulses require a separate setup and require the pulse to pass through additional dispersive material that may not be present in the SSSI setup. Conventionally, the probe spectral phase is determined by tracking the location (in an imaging spectrometer) of the pump-induced phase shift on the probe as the pump-probe delay is varied \cite{Wahlstrand:09}. This method can be time-consuming when using probe pulses with a very large bandwidth or when using low-repetition-rate laser sources. 

Recently, Vu \textit{et al.} have proposed an algorithm which may be used to circumvent this scanning procedure in favor of taking only a few pump-probe time-delayed shots \cite{Vu:10}. In this paper, we use the algorithm to retrieve the pump-induced probe phase shift in the time-domain using as few as two shots. We also propose and demonstrate a new scheme which requires only a single pump-probe shot to retrieve the modulation.

\section*{\normalsize Background}

In SSSI, two SC pulses, a reference pulse and a probe pulse (simply called the probe) upon which a time-delayed pump-induced phase shift has been imposed, interfere in the frequency domain in an imaging spectrometer. The reference pulse $E_{ref}(t)$ is often taken to be a replica of the probe which precedes the probe in time. As the probe pulse $E(t)$ co-propagates with the pump pulse (simply called the pump) through the medium that is being studied, it acquires a time-dependent phase shift $\Delta\Phi(t-\tau)$, where $\tau$ is the pump-probe delay. The resulting equation for the perturbed probe $\overline{E}(t) = E(t)e^{i\Delta\Phi(t-\tau)}$ can be solved to yield the time-domain phase shift occurring at $t=0$ as 

\begin{equation}\label{eq1}
    \Delta\Phi(t) = -i\ln\bigg\{\frac{F\{|\overline{E}(\omega)|e^{i\Delta\phi(\omega)+i\phi_s(\omega)-i\omega\tau} \}}{F\{|E(\omega)|e^{i\phi_s(\omega)-i\omega\tau} \}} \bigg\},
\end{equation}
where $F\{\}$ denotes the Fourier transform with respect to frequency, $|\overline{E}(\omega)|$ and $|E(\omega)|$ are the perturbed and unperturbed probe spectrums, respectively, $\Delta\phi(\omega)$ is the spectral phase shift of the probe induced by the pump, and $\phi_s(\omega)$ is the spectral phase of the unperturbed probe (same as that of the reference pulse). Of these quantities, $|\overline{E}(\omega)|$ and $|E(\omega)|$ can be obtained from the power spectrum of the probe and reference interference with and without the pump-induced modulation, respectively.

The spectral phase shift, $\Delta\phi(\omega)$, can be directly determined from the interferogram using a well-known Fourier transform technique \cite{Takeda:11}. The spectral phase of the probe, $\phi_s(\omega)$, is conventionally determined by studying the probe’s differential power spectrum, $\Delta I(\omega) = |E(\omega)|^2 - |\overline{E}(\omega)|^2$, as a function of the pump-probe time delay \cite{Tokunaga:12}. The spectral phase can be expanded about a central wavelength $\omega_c$ as

\begin{equation}\label{eq2}
 \phi_s = \phi_0 + b_1(\omega-\omega_c) + b_2(\omega-\omega_c)^2 + b_3(\omega-\omega_c)^3 + ...,
\end{equation}
where $\phi_0$ is the absolute phase, $b_1$ is related to the pulse shift in time, $b_2$ and $b_3$ are the second and third order dispersion coefficients, respectively. By tracking the location of a central minimum of the pump-induced phase shift on the probe in the frequency domain as the pump-probe time-delay is changed, a polynomial fit can extract $b_n$ to arbitrary order $n$ \cite{Wahlstrand:09}.  This method (hereafter referred to as the “conventional method” or “scan method”) requires one to scan the pump induced modulation across the entire spectrum of the probe pulse.

Vu \textit{et al.} presented a method that used a genetic algorithm (GA) to simultaneously characterize the probe spectral phase, that is, determine $b_i$ for $i=2,3,$…, and retrieve the pump-induced probe phase shift using as few as 2 time-delayed shots \cite{Vu:10}. The GA uses an initial population of $b_i$, and experimental interferograms (at different $\tau$ pump-probe delays) to retrieve a set of $\Delta\Phi(t)_{\tau}$. Notice that the factor $e^{-i\omega\tau}$ in Eq. (\ref{eq1}) shifts a modulation occurring at $t = \tau$ to $t=0$. This allows the algorithm to compare the shapes of the modulations from different time-delayed shots for trial $b_i$ and produce a corresponding fitness value using a cost function \cite{Vu:10},

\begin{equation}\label{eq3}
\Delta S^2 = \sum_{\tau}\int_{-\infty}^{\infty}[\Delta\Phi(t)_{\tau} - \overline{\Delta\Phi(t)}]^2 dt,
\end{equation}
where $\overline{\Delta\Phi(t)}$ is the average of the different $\Delta\Phi(t)_{\tau}$. Based on this comparison, the current generation of $b_i$ is modified to produce the next generation. This process is iterated until the algorithm minimizes the cost function. The proof showing that there exists a unique minimum of the cost function, in the stationary phase approximation, is given in \cite{Vu:10}. Here, we show an experimental demonstration of this algorithm and also present a new method to allow retrieval using only a single time-delayed shot instead of the minimum of 2 time-delayed shots that was originally proposed by Vu \textit{et al}.

\begin{figure}[t]
\centering\includegraphics[width=12cm]{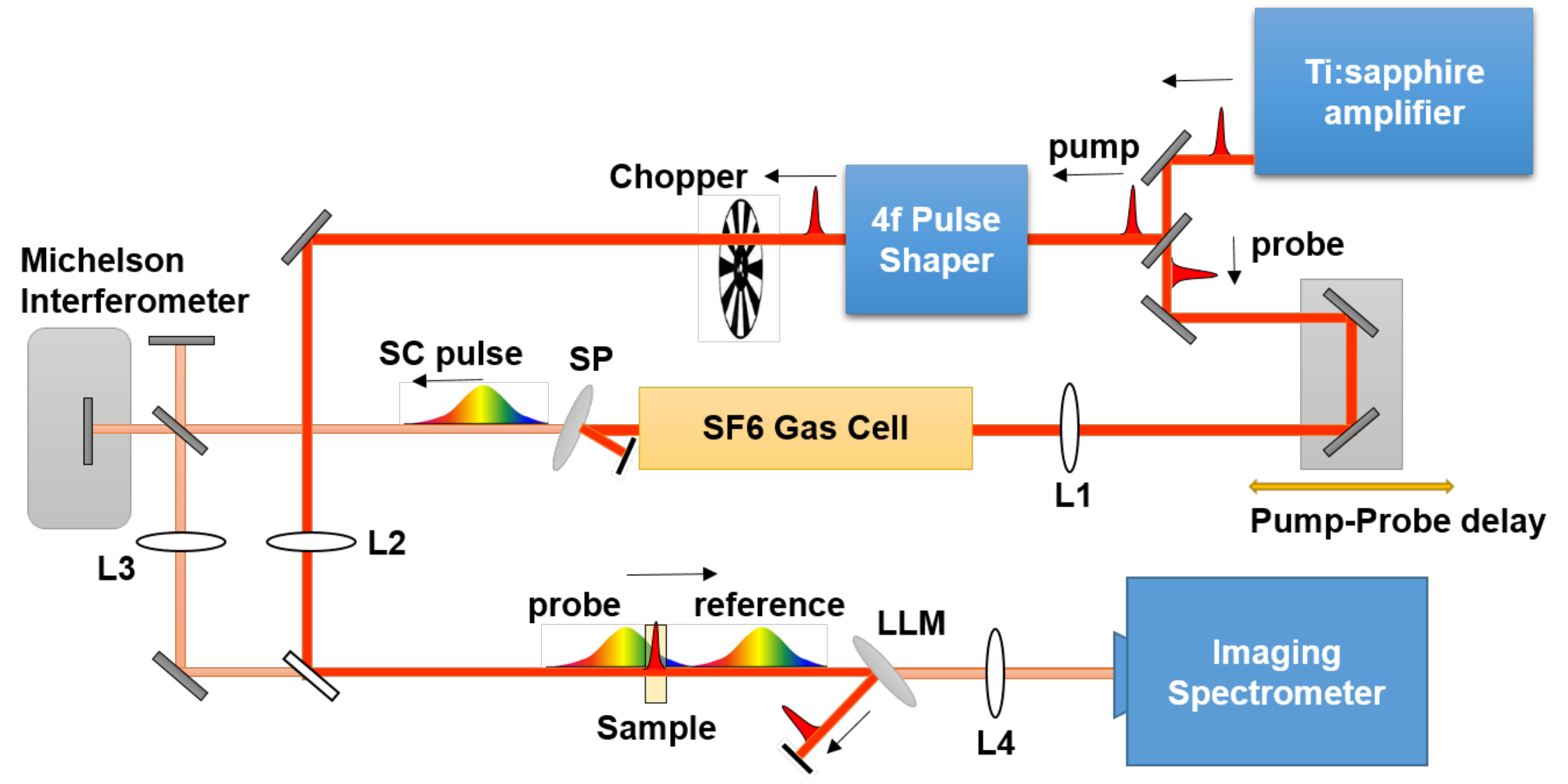}
\caption{Schematic diagram of the experimental setup. A pulse from a Ti:Sapphire amplifier is split into a probe pulse and a pump pulse. The probe undergoes filamentation in a SF6 gas cell and forms a supercontinuum (SC) pulse. It is further split into a reference pulse and a probe pulse in a Michelson interferometer. The pump modulates the probe in a sample and is then discarded by a laser line mirror (LLM). The reference and probe pulses interfere in an imaging spectrometer. SP is a shortpass filter. L1-L4 are lenses.}
\end{figure}

\section*{\normalsize Experimental Setup}

A schematic diagram of the experimental setup is shown in Fig. 1. Laser pulses of approximately 40 fs duration centered at 800 nm from a 1 kHz Ti:sapphire regenerative amplifier are split into two arms using a 80/20 (R/T) beamsplitter. The reflected arm, modulated by an optical chopper at 500 Hz, provides the pump beam to study nonlinearities in samples. A pulse shaper was used to shape the pump pulse in the second example studied, but not the first. The second (transmitted) arm, which provides the probe beam, is propagated through a gas cell containing SF6 gas at 1.9 bar to generate a 400-700 nm supercontinuum (SC) pulse via filamentation. The SC pulse is then split into a reference pulse and a probe pulse by a 50/50 beamsplitter in a Michelson interferometer. The reference pulse precedes the probe pulse by about 1.5 ps. The pump beam, focused by a lens into the sample, is collinearly combined with the probe beam such that the probe pulse is spatially and temporally overlapping with the pump as both propagate through the sample. Upon exiting the sample, the pump beam is rejected by a dichroic mirror and the probe and reference pulses are relay-imaged onto the entrance slit of an imaging spectrometer. The two pulses interfere, in the frequency domain, inside the spectrometer, and the resulting 2D spectral interferogram is imaged onto a CCD camera. In this study, we tested the algorithm by feeding it single-shot pump-probe interaction data, as well as data obtained from averaging 50 frames (at a fixed pump-probe time-delay) for improved signal-to-noise ratio (SNR). Both results will be shown in the examples below.

\begin{figure}[t]
\centering\includegraphics[width=10cm]{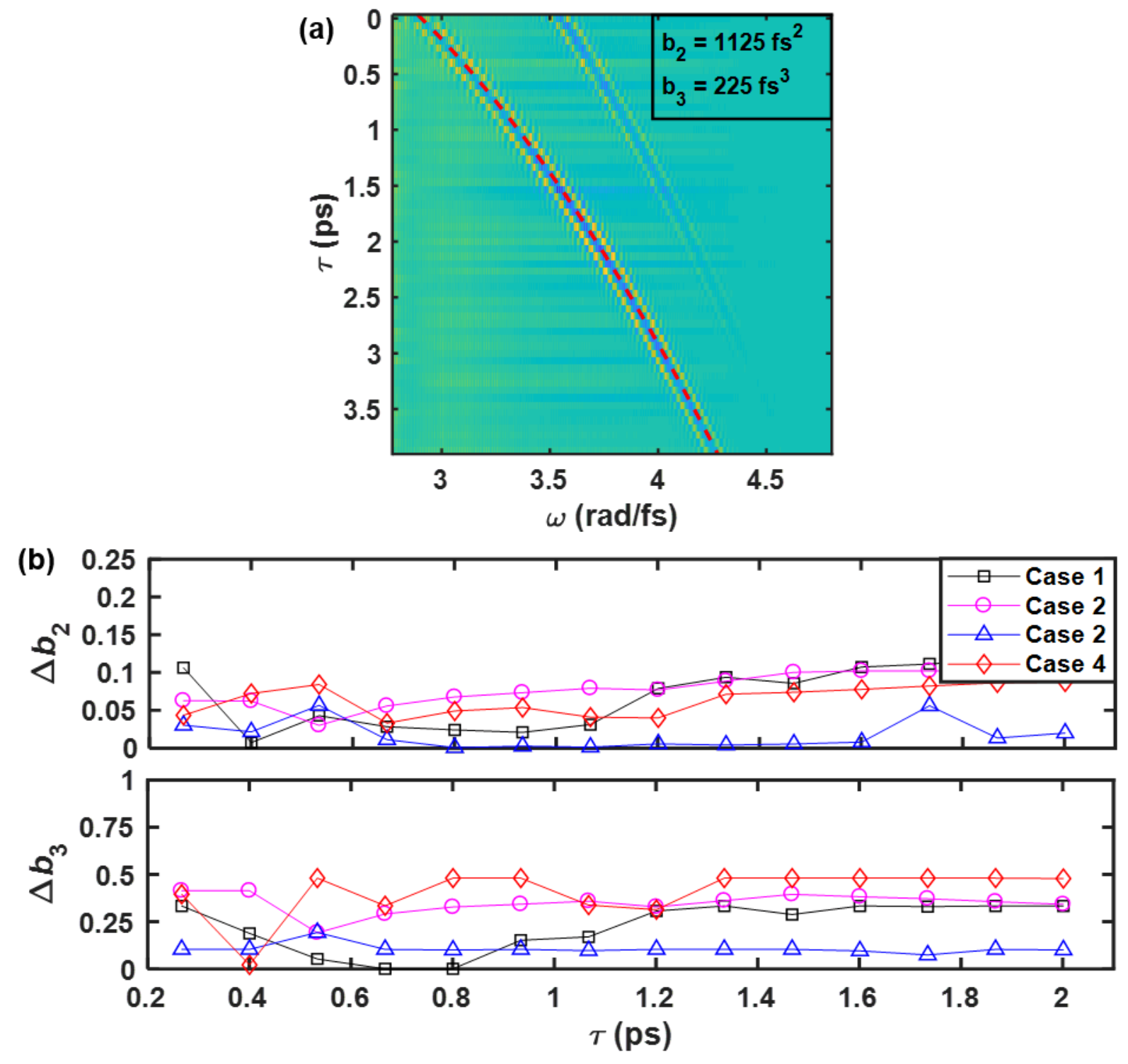}
\caption{(a) Probe pulse differential power spectrum as function of frequency (x-axis) and pump-probe time-delay (y-axis). The dotted red line is the fitted third-order polynomial used to extract $b_{2,3}$. (b) Top (Bottom): Fractional error in GA-retrieved $b_2 (b_3)$, $\Delta b_2 (\Delta b_3)$, for various pump-probe time-delays, $\tau$, for 4 cases each with different probe $b_{2,3}$. For Case 1,  $\overline{b}_2=1125$ fs\textsuperscript{2} and $\overline{b}_3=225$ fs\textsuperscript{3}. For Case 2, $\overline{b}_2=1361$ fs\textsuperscript{2} and $\overline{b}_3=256$ fs\textsuperscript{3}. For Case 3, $\overline{b}_2=1515$ fs\textsuperscript{2} and $\overline{b}_3=502$ fs\textsuperscript{3}. For Case 4, $\overline{b}_2=1747$ fs\textsuperscript{2} and $\overline{b}_3=289$ fs\textsuperscript{3}. Here $\Delta b_i \equiv |\overline{b}_i - b_i|/\overline{b}_i$.}
\end{figure}

\section*{\normalsize Examination of pump-probe delay dependence}
As a simple case, we studied the cross-phase modulation of the probe induced by the pump in a 500 $\mu$m thick plate of fused silica. Figure 2(a) shows the lineouts of the probe differential power spectrum, $\Delta I(\omega)$, for a set of time-delayed pump-probe shots where each shot is averaged over 50 frames. These were used to perform a polynomial fit to extract $b_{2,3}$ in the conventional way, which we will call $\overline{b}_{2,3}$.

Before we could test the GA, it was important to determine whether there existed a minimum temporal spacing between any two shots such that for temporal separations shorter than this minimum, the GA would not retrieve the probe spectral phase effectively. Using a stationary phase approximation for the Fourier transform of the time domain perturbed probe pulse, Vu \textit{et al.} proposed a minimum temporal separation criterion between two time-delayed shots $\Delta\tau_{min} > (\epsilon b_2^2)/(6b_3)$ where $\epsilon$ is the measurement error associated with $b_3$ \cite{Vu:10}. This condition can be thought of as a direct consequence of the $b_2$ term dominating the local variation in the spectral phase of a chirped pulse that has insignificant higher order spectral phase, $b_n$ for $n>$ 3. If the temporal separation between two shots becomes too small then the GA would not be able to accurately determine the contribution of $b_3$, resulting in an inaccurate probe spectral phase retrieval.

To test this condition, we conducted pump-probe measurements (via the full scanning method) in the fused silica for four different cases. In each case the probe/reference pulse spectral phase was altered by adding or removing some piece of dispersive material in the probe beam path. In the first case (Case 1), no dispersive material was added to the probe beam path. In the second (Case 2), third (Case 3) and fourth (Case 4) experiments, an uncoated 6.35--mm thick BK7 glass, a 0.5--mm thick TiO\textsubscript{2}, and a 2--cm thick fused silica, respectively, were added to the probe beam path.  For each of the experiments, the GA was used to retrieve $b_2$ and $b_3$ from pairs of shots with a variable time delay (each shot averaged over 50 frames). Figure 2(b) shows the normalized deviation of these retrieved coefficients, $\Delta b_i \equiv |\overline{b}_i - b_i|/\overline{b}_i$, where $i=2,3$, from $\overline{b}_{2,3}$ (the conventionally retrieved coefficients).

We see that the GA-retrieved probe spectral phase did not improve consistently with increasing time delay between two shots for any of the cases. Thus, we were not able to verify the condition on minimum temporal separation between shots. This may be attributed to the \say{allowable} error in the spectral phase of the probe that is used in the retrieval. To avoid distortion in the shape of the retrieved modulation, the dispersion coefficients must satisfy the relation $|\overline{b}_i - b_i| < (\tau_\omega/\pi)^i$, where $\tau_\omega$ is the temporal duration of the pump-induced modulation on the probe and $\overline{b}_i$ is the true value of the coefficient, here taken to be that which is retrieved via the conventional method \cite{Wahlstrand:09}. This is true for a given coefficient $b_n$ if all other coefficients are determined exactly. For our case with no dispersion added to the probe beam path, $\tau_\omega \approx 40$ fs, $b_2=1125$ fs\textsuperscript{2}, and $b_3=225$ fs\textsuperscript{3}. So we easily calculate that if  $b_2$ is retrieved by the GA to be $b_2= 1125 \pm 113$ fs\textsuperscript{2} (i.e. $\Delta b_2 = 0.01$) then any $b_3 \approx 225 \pm 625$ fs\textsuperscript{3} will result in a time-domain retrieval of the probe phase shift that is not significantly distorted as compared to the true modulation. Naturally the bounds are tighter on $b_2$ than $b_3$, and this is reflected in Fig. 2(c)(bottom), which shows that the GA determines $b_2$ more accurately than $b_3$.

\begin{figure}[t]
\centering\includegraphics[width=10cm]{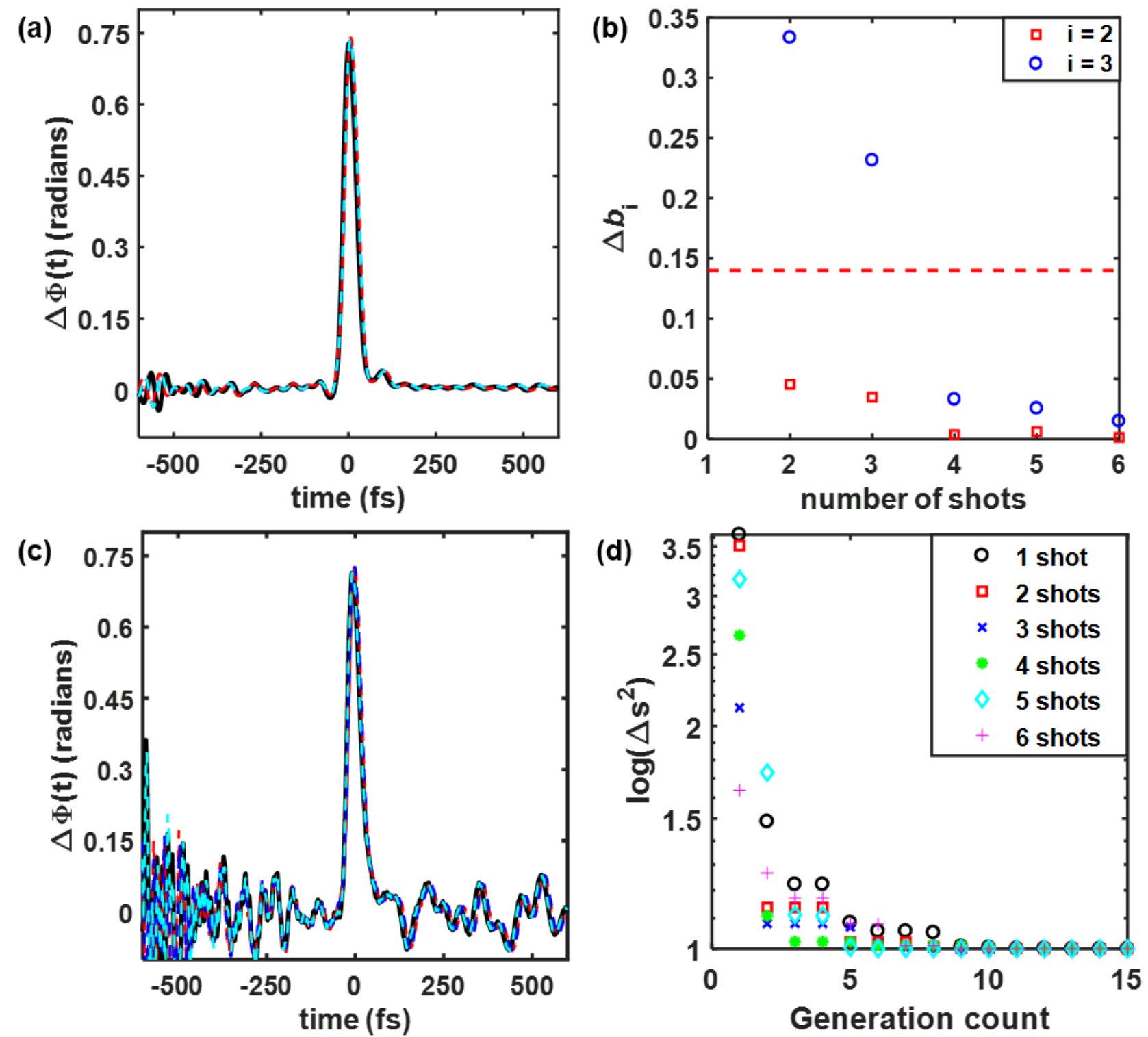}
\caption{(a) Experimental comparison between conventionally-retrieved (solid black line) and GA-retrieved modulation with each measurement averaged over 50 frames (for the 2,3 and 4 shots cases). 2-shots case (dashed red line): $b_2=1176$ fs\textsuperscript{2}, $b_3=150$ fs\textsuperscript{3}. 3-shots case (dashed blue line): $b_2=1164$ fs\textsuperscript{2}, $b_3=173$ fs\textsuperscript{3}. 4-shots case (dashed cyan line): $b_2=1129$ fs\textsuperscript{2}, $b_3=217$ fs\textsuperscript{3}. (b) Fractional error in GA-retrieved $b_{2,3}$ as a function of shots used. Red squares correspond to $\Delta b_2$ and blue circles are values for $\Delta b_3$. Dashed red line represents $(\Delta b_2)_{max}$ that is allowed beyond which retrieved modulation becomes significantly distorted. (c) Same as in (a) but without averaging over multiple frames. 2-shots case: $b_2=1173$ fs\textsuperscript{2}, $b_3=150$ fs\textsuperscript{3}. 3-shots case: $b_2=1167$ fs\textsuperscript{2}, $b_3=168$ fs\textsuperscript{3}. 4-shots case: $b_2=1129$ fs\textsuperscript{2}, $b_3=217$ fs\textsuperscript{3}. (d) $\log(\Delta s^2)$ vs generation count. $\Delta S^2$ is normalized to its final converged value.}
\end{figure}

\section*{\normalsize Kerr-induced nonlinearity in fused silica}
Using a temporal separation of  $\sim$600 fs between shots, we used the GA to retrieve $b_{2,3}$ with up to 6 time-delayed shots. The conventionally retrieved time-domain probe phase shift is compared with the GA-retrieved ones for the 2, 3, and 4 shots cases in Fig. 3(a) and Fig. 3(c) using 50-frame averaged and single-frame data, respectively. We can see that the peak phase shift incurred by the probe beam is $\Delta\Phi(t)_{peak} \approx 0.73$. We knew the pump pulse was centered at 800 nm, had energy of about 2.5 $\mu$J and a duration of about 40 fs. Using $n_2= 4.3 \times10^{-20}$  m\textsuperscript{2}/W for fused silica \cite{Cimek:13} we can estimate a peak phase shift of the probe due to cross-phase modulation as $\Delta\Phi(t)_{peak} = 2k_0Ln_2I \approx 0.75$, where $k_0$ and $I$ are the free space wavenumber and optical intensity of the pump pulse, and $L$ is the thickness of the fused silica plate. The estimated value agrees well with the retrieval. The modulation duration is approximately 40 fs, which also agrees very well with the pump pulse duration. We see that although the SNR in Fig. 3(c) is noticeably worse than in Fig. 3(a) (i.e. a single frame data vs. 50-frame averaged data), both the retrieved probe phase coefficients and the time-domain modulation characteristics are still in excellent agreement.

In Fig. 3(b) the error in the retrieved coefficients, $\Delta b_i$, is seen to decrease with an increasing number of shots. For the nonlinear phase shift experienced by the probe pulse in the fused silica, $\tau_\omega$ was about 40 fs which means the maximum allowable deviation $(\Delta b_2)_{max} \approx 0.14$ (dashed red line in Fig. 3(b)) and $(\Delta b_3)_{max} \approx 9.17$ (not shown in Fig. 3(b)). Even with just 2 shots, the GA retrieves both coefficients within these bounds so as to not significantly distort the shape/duration of the modulation. With 4 shots, the GA error falls well below 0.5\% for $b_2$ and 5\% for $b_3$.  Figure 3(d) shows a plot of the $\Delta s^2$ values for each generation for the 2, 3, 4, 5 and 6 shots cases. Here, $\Delta s^2$ are obtained by normalizing the $\Delta S^2$ values to the final converged value. We see that the algorithm converges faster with increasing number of shots. In fact, even with just 2 shots it converges within 15 generations.

\begin{figure}[t]
\centering\includegraphics[width=10cm]{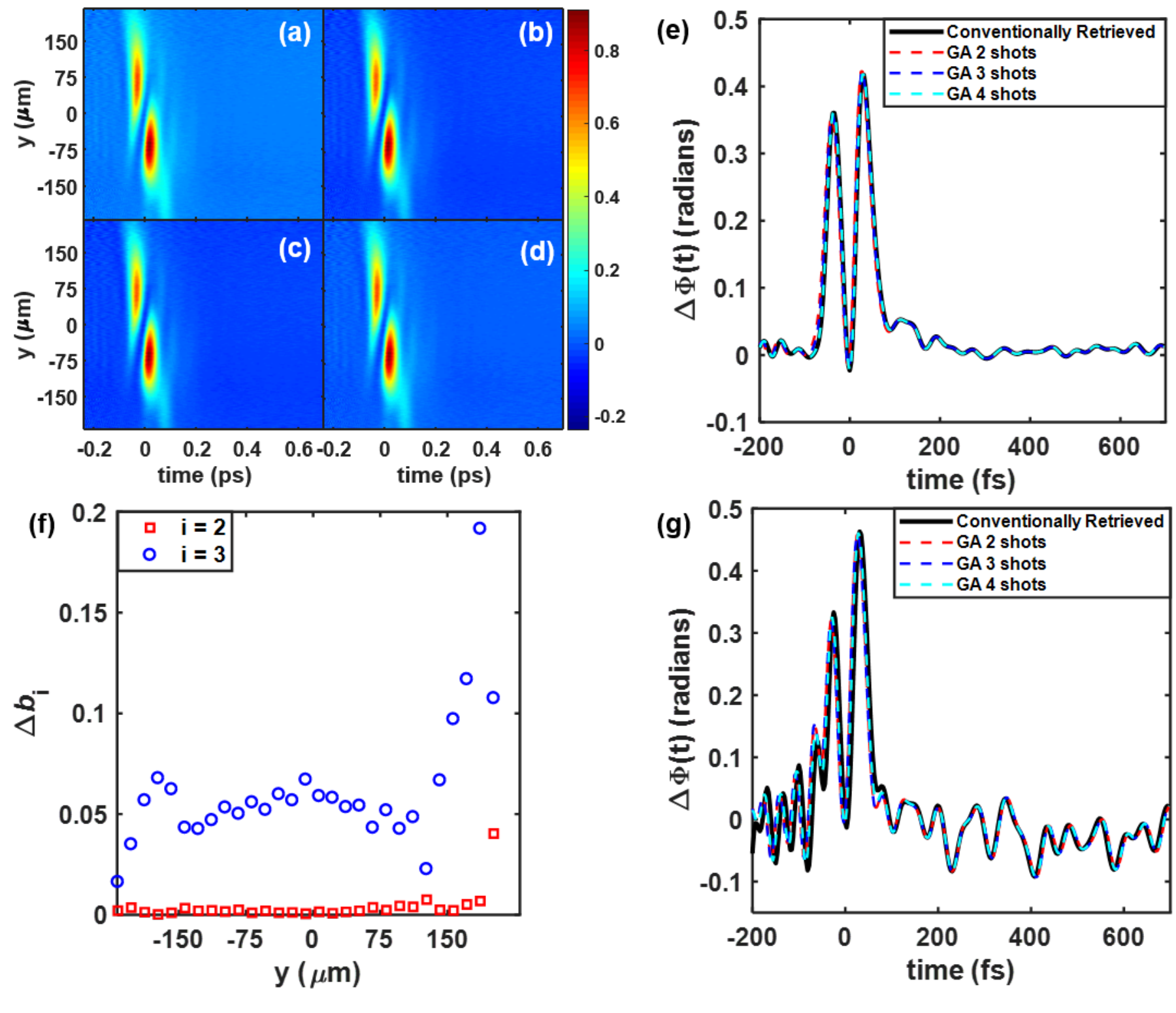}
\caption{Far-field image of STOV-carrying pump pulse-induced phase shift on probe. The data in (a)-(f) are obtained from measurements that were averaged over 50 frames at a given pump-probe delay; (g) represents data obtained without averaging over multiple frames. (a) Conventionally retrieved: $b_2=1144$ fs\textsuperscript{2}, $b_3=215$ fs\textsuperscript{3}. (b) 2 shots GA-retrieved: $b_2=1153$ fs\textsuperscript{2}, $b_3=165$ fs\textsuperscript{3}. (c) 3 shots GA-retrieved: $b_2=1150$ fs\textsuperscript{2}, $b_3=184$ fs\textsuperscript{3}. (d) 4 shots GA-retrieved: $b_2=1141$ fs\textsuperscript{2}, $b_3=204$ fs\textsuperscript{3}. (e) Lineouts at location $y=0$ from (a)-(d). Comparison between the conventionally retrieved modulation (solid black line) and GA-retrieved using 2 (dashed red line), 3 (dashed blue line), and 4 (dashed cyan line) shots. (f) Fractional error in GA-retrieved $b_{2,3}$ at different points along slit dimension. Red squares correspond to $\Delta b_2$ values and blue circles to $\Delta b_3$ values. (g) Experimental comparison between conventionally retrieved and GA-retrieved modulations from single frame data. 2 shots GA: $b_2=1150$ fs\textsuperscript{2}, $b_3=184$ fs\textsuperscript{3}. 3 shots GA: $b_2=1150$ fs\textsuperscript{2}, $b_3=176$ fs\textsuperscript{3}. 4 shots GA: $b_2=1142$ fs\textsuperscript{2}, $b_3=201$ fs\textsuperscript{3}.}
\end{figure}

\section*{\normalsize Imaging of spatio-temporal optical vortex pulses}
Next, we studied the probe phase shift induced by a spatio-temporal optical vortex (STOV)-carrying pump pulse. STOVs are optical structures that arise naturally during pulse collapse and arrest of self-focused beams due to transient phase shear resulting in a phase discontinuity \cite{Jhajj:14}. STOVs can also be artificially generated by applying spatiotemporal index transients to a pulse \cite{Hancock:15}. A 4f pulse shaper, as described in \cite{Hancock:15}, was used to generate such pump pulses. A reflective grating was used to Fourier transform the original pump pulse into the frequency domain and then a $\pi$-step phase-mask was used to apply a field null to it. The pump pulse was then reverted back to the time domain using another reflective grating. At the output of the pulse shaper a “doughnut shaped” STOV is formed and at the focus of a lens, the far-field projection of a STOV in free space propagation is formed. We used such a pump pulse to modulate the probe in a 500 $\mu$m thick fused silica plate, resulting in a spatially- and spectrally- complicated modulation. The full 2D (spatial and spectral) pump-induced probe phase shift (averaged over 50 frames at a fixed pump-probe delay) was retrieved using both the conventional method and the GA (for 2,3 and 4 shots cases).

The results are shown in Figs. 4(a-d). These are the far-field images of a “line-STOV”, and so what we see is not the characteristic donut shape of the STOV but rather two lobes offset in space and time. Despite the complicated spatial and spectral structure of this modulation, the GA was able to retrieve $b_2$ and $b_3$ within 0.5\%  and 5\%, respectively, of the conventionally retrieved values using 4 shots. The GA retrieved modulation (for the 2, 3, and 4 shots case) is in excellent agreement with the modulation retrieved conventionally. A lineout from each of the retrievals is shown in Fig. 4(e). For comparison, we also show the retrieved time-domain modulation extracted from a single frame data in Fig. 4(g). Although the time-domain modulation from the single frame data is noisy, the spectral phase coefficients of the probe are still determined accurately.

Moreover, we used the GA to retrieve $b_{2,3}$ coefficients for several positions along the spatial dimension $y$. The beam used in this experiment had no known spatial chirp, and this is reflected in the consistency of the GA-retrieved coefficients at different spatial locations as shown in Fig. 4(f). The largest deviation of $b_2$ and $b_3$ is less than 1\% and 20\%, respectively. The $\sim$20\% $b_3$ deviation may seem large but we must recall that given only a 1\% deviation in $b_2$, $b_3$ may deviate from the true value by almost as much as 280\% before introducing significant distortion in the retrieved temporal modulation. As in this example, a lineout from any location on the interferogram can be used to determine any present spatiotemporal chirp in the probe even when the modulation has features that vary strongly in the spatial and spectral domains.

\section*{\normalsize Consideration of higher order  dispersion}
It was shown in \cite{Wahlstrand:09} that extending the polynomial fit of the probe differential power spectrum as a function of pump-probe delay to include higher-order dispersion coefficients ($b_n$, $n>3$) could result in a more reliable time-domain retrieval of the probe phase shift. We modified the GA to retrieve higher-order $b_n$ (results not shown) possibly present in the previous examples and found that the retrieved modulation was not particularly improved compared to keeping only the $b_2$ and $b_3$ terms. In particular, the small oscillations on either sides of the modulation peak, potentially from imperfect probe spectral phase characterization, were not damped as higher-order $b_n$ were used in the retrieval. Therefore, we concluded that for well-behaved probe pulses with mostly second- and third- order spectral phase, retrieving the higher-order coefficients using the GA added no practical benefit.

\section*{\normalsize Retrieval from a single time-delayed shot}
So far we have used the GA with at least two time-delayed shots. Surprisingly, our scheme can work with only a single shot when the reference and probe pulses are identical and partially overlapping in time. If a pump-induced phase shift occurs in the temporal region where the reference and probe pulses overlap, then it will induce a phase-shift in different parts of the probe spectrum and reference spectrum. Since the probe and reference pulses are identical, the role of the reference and probes can be switched. We may treat the single modulation on the reference and the probe as two $\tau_r$-delayed modulations on a single probe (as shown in Fig. 5(a)), where $\tau_r$ is the temporal separation between the reference and the probe pulses. Since the spectral resolution, which sets a limit on the ultimate achievable temporal resolution \cite{Wahlstrand:09}, is determined, via the Nyquist frequency, to be twice the fringe spacing, $\Delta\omega = 2\omega_s = 4\pi/\tau_r$ where $\omega_s$ is the fringe spacing. Here we would think to maximize $\tau_r$ to the maximum allowable by our spectrometer resolution. Doing this, however, would reduce the temporal overlap of the reference and probe, thus reducing the temporal window that can be used to observe the pump-induced modulation. This problem can be mitigated by using probe/reference pulses that are highly chirped and therefore longer in duration. Also, the probe/reference overlap in the medium we are studying should not cause any issues since both are very weak and will not interact nonlinearly. In addition, we extract the probe phase shift by subtracting the spectral phase of the probe/reference interference with and without the pump, so any potential nonlinear interactions between the probe and reference due to their overlap will become irrelevant in the final retrieval.

\begin{figure}[t]
\centering\includegraphics[width=10cm]{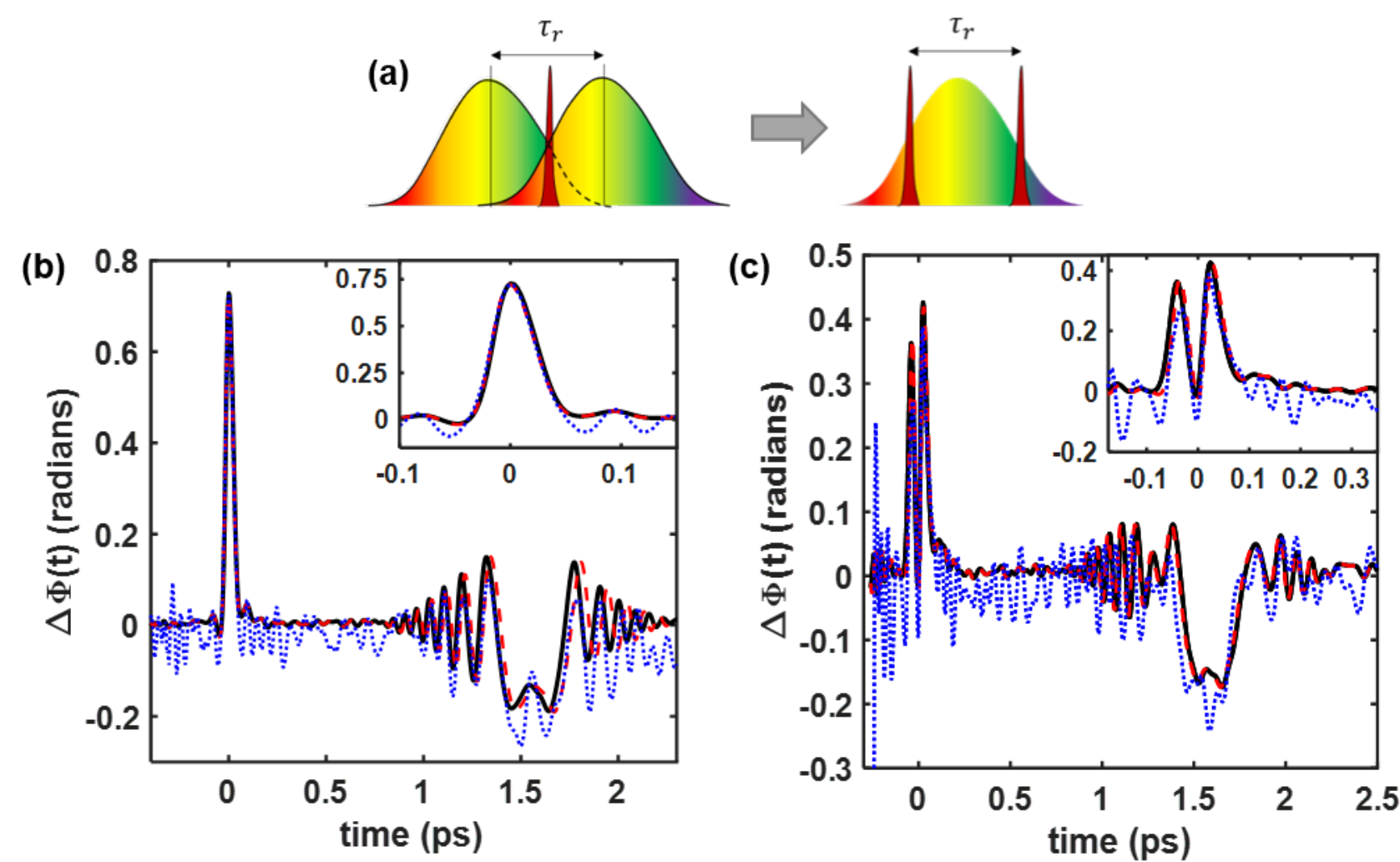}
\caption{(a) A pump-induced modulation on the reference and probe in a region where the two overlap is equivalent to two identical modulations applied on a single probe, separated in time by the reference-probe separation. (b) The solid black line is the conventionally retrieved modulation. The dashed red line is the single-shot, GA-retrieved modulation from a measurement averaged over 50 frames: $b_2=1108$ fs\textsuperscript{2}, $b_3=268$ fs\textsuperscript{3}. The dotted blue line is the single-shot, GA-retrieved modulation from a single frame measurement: $b_2=1145$ fs\textsuperscript{2}, $b_3=242$ fs\textsuperscript{3}. The inset shows a zoomed-in view of the pump-induced phase shift on the probe. (c) Single-shot, GA-retrieved modulations averaged over 50 frames ($b_2=1120$ fs\textsuperscript{2}, $b_3=235$ fs\textsuperscript{3}) and from a single frame measurement ($b_2=1084$ fs\textsuperscript{2}, $b_3=325$ fs\textsuperscript{3}). Same color scheme as in (b).}
\end{figure}

Note that there are two curves in the probe differential power spectrum in Fig. 2(a); one is the pump-induced modulation on the probe (which was fitted to obtain the dispersion coefficients) and the other is the pump-induced modulation on the reference. The retrieved time-domain modulation on the probe (reference) will consist of the modulation and an “echo” – the pump-induced modulation on the reference (probe) part of the spectrum. The presence of the “echo” at a fixed time delay, $\tau_r$, will serve as an additional constraint for the GA – resulting in a more accurate retrieval. As mentioned before, the factor $e^{-i\omega\tau}$ in Eq. (\ref{eq1}) will shift a modulation occurring at $t=\tau$ to $t=0$. This means that if the GA retrieves the modulation using incorrect spectral phase coefficients, the temporal spacing between the modulation and echo and/or their shapes will be distorted.

To demonstrate this single-shot method, we used the GA to retrieve the time-domain pump-induced modulation on the probe (and reference) from the above described experiment in fused silica, as well as for the experiment using a STOV-carrying pump pulse. The results are shown in Figs. 5(b) and 5(c). We see that $\Delta b_2 < 0.03$ and $\Delta b_3 < 0.20$ for both cases using this single-delayed-shot method (for the measurements obtained by averaging over 50 frames). In addition, the retrieved modulations are in good agreement with those retrieved with the conventional method. We should mention that, as seen in Fig. 5(c), the modulation retrieved from a single frame is somewhat noisy because the data used has a poor SNR and the modulation has a complicated structure.

\section*{Conclusion}
In conclusion, we have experimentally demonstrated the algorithm proposed by Vu \textit{et al} \cite{Vu:10}. It was shown that the algorithm is successful in retrieving spatially- and spectrally-complicated probe spectral phase shifts. The accuracy of the retrieved modulations was seen to improve dramatically with increasing number of shots. A single-shot variation was proposed and tested to yield highly accurate results. This variation only requires that the pump-induced phase shift be placed in a temporal region where the probe overlaps with the reference pulse. This technique would prove especially useful in conducting SSSI using a low-repetition-rate laser source. 

\section*{Funding}
Office of Naval Research (N00014-17-1-2705); Air Force Office of Scientific Research (FA9550-16-1-0163); National Science Foundation (NSF) (1351455)

\end{document}